\documentclass [11 pt]{extarticle}
\usepackage[utf8]{inputenc}
\usepackage{bm}
\usepackage{amsthm}
\usepackage{hyperref}

\usepackage{soul}
\usepackage{graphicx}
\usepackage{mathtools}
\usepackage{comment}
\usepackage{amsmath}
\usepackage{xcolor}
\usepackage{upgreek}
\usepackage{amssymb}
\usepackage{subfigure}
\usepackage[makeroom]{cancel}
\usepackage{geometry}
\numberwithin{equation}{section}
\usepackage{cite}
\title{Managing spectral properties and Schmidt mode content of squeezed vacuum light using sum-frequency converter}

\usepackage{authblk}

\author[1,*]{Vladislav Sukharnikov}
\author[2]{Polina Sharapova}
\author[1,3]{Olga Tikhonova}

\affil[1]{Faculty of Physics, Lomonosov Moscow State University, Leninskiye Gory 1, Moscow 119991, Russian Federation}
\affil[2]{Department of Physics, University of Paderborn, Warburger Stra{\ss}e 100, Paderborn D-33098, Germany}
\affil[3]{Skobeltsyn Institute of Nuclear Physics, Lomonosov Moscow State University, Moscow 119234, Russian Federation}

\affil[*]{\textit{Contact e-mail sukharnikov.vladislav@gmail.com}}
\date{}

\begin{document}

\maketitle

\begin{abstract}
	Capabilities of quantum optical SFG-gate seeded by squeezed light are investigated in the frame of frequency Schmidt modes. Methods to manage and manipulate extensively the properties and mode content of squeezed light are developed. Possibilities to block and select any certain Schmidt mode of squeezed light with conservation of non-classical properties are demonstrated. The significant phase sensitivity of the gate is shown and the ways to manage the spectral distribution of the output light due to the phase effects and variable coupling between modes in the gate are demonstrated. The effect of swapping between modes in the gate is found. It allows to enhance squeezed light in a set of modes without loss of photon correlations which is important for further experiments and new applications. 
\end{abstract}

\section{Introduction}

Nowadays non-classical squeezed states of light become extremely attractive for researchers from many fields of physics due to their unique features. Suppressed fluctuations for one of the field quadratures of squeezed vacuum light is a crucial step towards the noise reduction beyond the standard quantum limit which is strongly required for high resolution measurements 
\cite{abadie2011gravitational, aasi2013enhanced, grote2013first, dooley2015phase, oelker2016ultra} and quantum imaging tasks \cite{brida2010experimental, lopaeva2013experimental}. Squeezed light can be generated in the nonlinear optical processes, for instance in the parametric down-conversion process, which produces the squeezed vacuum state or twin-beams of two mode squeezed light \cite{Agafonov:2010}. In high-gain regime the generated squeezed light can be referred to as a macroscopic quantum state, since it may contain up to $10^{13}$ photons \cite{iskhakov2012superbunched}, which exhibit strong entanglement  \cite{stobinska2012entanglement, iskhakov2012polarization, chekhova2015bright}, two-mode squeezing \cite{Iskhakov:2009, kalashnikov2012measurement} and correlations in both spectral  \cite{Spasibko:2012} and spatial \cite{sharapova2015schmidt} domains.

The analytical description of squeezed vacuum light is usually performed by the introduction of new broadband operators describing photons in the so-called Schmidt modes \cite{wasilewski2006pulsed, Fedorov:2009, christ2013theory}. This approach allows one to take into account each mode independently, as it was done in a spectral analysis in \cite{Sharapova:2018}.
Due to vastly multimode structure of squeezed vacuum light it proves to be a useful tool for quantum information science, because temporal modes may be used for encoding of quantum information \cite{brecht2015photon, ansari2018tailoring}.

It is a non-trivial task to manipulate and fully utilize all features possessed by the squeezed vacuum light, since quantum properties of squeezed light can be easily ruined by the losses and noises brought by usually used optical devices, thus the quantum-optical methods are strongly required. For example, in \cite{lemieux2016engineering} engineering of frequency spectrum has been conducted by the application of dispersive medium. However, more successfully such tasks are solved by the application of scheme based on the sum-frequency generation (SFG) process seeded by squeezed vacuum light \cite{reddy2017engineering, shahverdi2017quantum}, which were shown to be able to block a selected temporal mode  \cite{brecht2011quantum, eckstein2011quantum, manurkar2016multidimensional, allgaier2018pulse} and to solve the quantum tomography tasks \cite{ansari2017temporal, ansari2018tomography}. Such scheme is called quantum pulse gate or quantum optical gate.

In this paper we describe theoretically the capabilities of quantum optical SFG-gate seeded by squeezed light in the frame of independent frequency Schmidt modes. Methods to manage and manipulate extensively the properties and the mode content of squeezed vacuum light are developed. We focus mostly on the multi-mode regime, in which signal mode for SFG has projections on two Schmidt modes of squeezed vacuum seed. In this case the interplay between initially independent Schmidt modes becomes possible, which may significantly affect the mode distribution and spectral properties of incoming squeezed vacuum light. Under such conditions, the photon correlations and degree of squeezing of the converted light are analyzed.

\section{Theoretical approach}
The starting point of our research is the description of sum-frequency generation process (SFG) in the formalism of Schmidt modes and corresponding Schmidt operators. Sum-frequency generation process is a second-order nonlinear optical process, in which two incoming signal and pump photons give rise to output photon of sum-frequency. The Hamiltonian of such process is given by:
\begin{equation}\label{eq:1}
\widehat{H}_\text{SFG} \sim \int d\mathbf{r}\, \chi^{(2)}\;  \widehat{E}_p^{(+)}\; \widehat{E}_{s}^{(+)} \;\widehat{E}_{o}^{(-)} + h.c.,
\end{equation}
where indices $p$, $s$ and $o$ correspond to pump, signal and output radiation fields respectively. Here $\chi^{(2)}(\mathbf{r})$ is the second-order susceptibility, which is further assumed to have a weak spatial dependence $\chi^{(2)}(\mathbf{r}) \approx \chi^{(2)}_0$. Integration runs over the whole crystal medium.

Further consideration is restricted to the collinear propagation of radiation fields, which allows to neglect the transverse components of wave-vectors, and to the degenerate regime of SFG, i.e. center frequencies of pump and signal fields match each other and are equal to $\omega_p$. The pump pulse is treated as classical field with intensity $E_0$ and normalized spectral distribution $\Phi(\omega-\omega_p)$ centered at $\omega_p$ with characteristic spectral width $\sigma$. Note that the pump and signal photons have an ordinary and extraordinary polarization respectively \cite{eckstein2011quantum}, which provides the distinction between these fields. 

In the considered SFG scheme signal and outgoing high-frequency photons are usually found to be strongly entangled in the spectral domain. However in some cases the regime without such correlations can be interesting and some methods to eliminate the entanglement are developed \cite{Grice:2001, eckstein2011quantum}. Appendix \ref{sec:Hamiltonian} shows that the entanglement between the signal and output photons may be suppressed under the condition that the pump duration is shorter than the characteristic time of delay between signal and output field due to dispersion:
\begin{equation}\label{eq:2}
1/\sigma \ll L\cdot \left|k_p'(\omega_p) - k_0'(2\omega_p)\right| ,
\end{equation}
where $L$ is the effective crystal length along the axis of propagation of fields. The fulfillment of this condition results in the factorization of the joint spectral amplitude into individual spectral modes for signal and outgoing photons, when the signal mode is determined only by the pump spectral envelope $\Phi(\omega)$, while the sum-frequency mode has well-defined Gaussian shape determined by the crystal parameters. The Hamiltonian \eqref{eq:1} rewritten in terms of new broadband creation and annihilation operators for these modes is:
\begin{equation}\label{eq:3}
\widehat{H}_\text{SFG} = i \hbar \Gamma ( \widehat{D}  \widehat{C}^\dagger - \widehat{D}^\dagger \widehat{C}),
\end{equation}
where $\Gamma \sim \chi^{(2)}\cdot E_0$ is the parametric gain (or effective coupling constant). The photon broadband creation operator $\widehat{C}^\dagger$ accounts for the creation of output photons of sum-frequency, with spectral distribution close to the Gaussian function:
\begin{equation}\label{eq:4}
\widehat{C}^\dagger = \sqrt[4]{\frac{1}{\pi \Delta \omega^2}} \int d\omega_o \, \exp\left\{-\frac{(\omega_o - 2\omega_p)^2}{2\Delta \omega^2}\right\}  \widehat{a}^\dagger_{\omega_o},
\end{equation}
where $1/\Delta \omega \sim L\cdot|k'_p(\omega_p) - k'_o(2\omega_p)|$ is the characteristic dispersion width playing role in condition \eqref{eq:2}. The broadband photon operator $\widehat{D}$ describes the annihilation of photons in signal mode:
\begin{equation}\label{eq:5}
\widehat{D} = \int d\omega_s \, \Phi(\omega_s - \omega_p) \, \widehat{a}_{\omega_s}.
\end{equation}

In this case the transformation of photon operators obtained as the solution of corresponding Heisenberg equations is given by:
\begin{equation}
\begin{pmatrix}\label{eq:6}
\widehat{C}^\text{out}\\
\widehat{D}^\text{out}
\end{pmatrix} = 
\begin{pmatrix}
\cos{\Theta} & \sin{\Theta} \\
-\sin{\Theta} & \cos{\Theta} 
\end{pmatrix}
\begin{pmatrix}
\widehat{C}^\text{in}\\
\widehat{D}^\text{in}
\end{pmatrix}.
\end{equation}
Such transformation coincides with the beamsplitter rotation matrix, with parameter $\Theta = \int dt\, \Gamma$ referred to as beamsplitter angle. Integration is taken over the characteristic time of interaction (i.e. pump pulse duration). Initial condition for sum-frequency mode is assumed to be vacuum, while the state for the signal mode is represented by some seeding field with non-zero mean number of photons. It is required for the non-zero output since the mean number of photons in the sum-frequency is the portion of initial number of photons in the signal mode:
\begin{equation}\label{eq:7}
\langle \widehat{N}^\text{out}_\text{SF}\rangle =  \sin^2 \Theta \cdot  \langle \widehat{N}^\text{in}_\text{s}\rangle,
\end{equation}
where $\widehat{N}_\text{s} = \widehat{D}^\dagger \widehat{D}$ and $\widehat{N}_\text{SF} = \widehat{C}^\dagger \widehat{C}$. The remaining portion of initial number of photons in signal mode is not converted and remains in the signal channel:
\begin{equation}\label{eq:8}
\langle  \widehat{N}^\text{out}_\text{s} \rangle = \cos^2 \Theta \cdot \langle \widehat{N}^\text{in}_\text{s} \rangle.
\end{equation}
Relations \eqref{eq:7} and \eqref{eq:8} reflect the conservation of total number of photons involved in the SFG-gate 
\begin{equation}\label{eq:9}
\widehat{N}_s^\text{out} + \widehat{N}_\text{SF}^\text{out} = \widehat{N}_s^\text{in}.
\end{equation}
From relation \eqref{eq:7} it follows that full conversion of signal mode correspond to the cosine zeros at beamplitter angles $\Theta = (k + 1/2) \pi$, where $k$ is integer. In this case all photons from signal mode are up-converted to sum-frequency.

We assume that the seed of SFG process is the squeezed vacuum light, generated in the degenerate parametric down-conversion process, which may be represented as the superposition of independent contributions of Schmidt modes. The corresponding state of seed is defined as the action of squeezing operator on vacuum \cite{Sharapova:2018}:
\begin{equation}\label{eq:10}
|\psi \rangle_s =  \exp{\left\{\dfrac{G}{2} \sum_{n=0}^{\infty} \sqrt{\lambda_n} \left( \widehat{A}^{\dagger 2}_n - \widehat{A}_n^2 \right) \right\}} |0 \rangle_s,
\end{equation}
where $G$ is the squeezing parameter and $\widehat{A}_n, \widehat{A}_n^\dagger$ are broadband annihilation and creation operators of photons in the corresponding Schmidt mode with spectral distribution $u_n(\omega)$. In a most simple case the Schmidt mode profiles coincide with normalized Hermite-Gaussian functions. Spectral distribution of squeezed vacuum is defined as the following double summation:
\begin{equation}\label{eq:11}
\mathcal{N}(\omega) = \sum_{n,m} u_n(\omega) \cdot u_m^*(\omega) \langle \widehat{A}_m^{\dagger}  \widehat{A}_n \rangle.
\end{equation}
For the independent Schmidt modes the normalized spectral distribution is defined as follows:
\begin{equation}\label{eq:12}
\mathcal{N}(\omega) = \sum_{n=0}^{\infty} |u_n(\omega)|^2 \cdot \Lambda_n.
\end{equation}
The weight of Schmide mode $\Lambda_n$ is defined as the ratio of mean number of photons in this mode to the total number of photons in squeezed vacuum. The mean value of photons in $n$-th mode is well-known \cite{sharapova2015schmidt}:
\begin{equation}\label{eq:13}
N_n = \langle \hat{N}_n \rangle = \sinh^2\left(G\sqrt{\lambda_n}\right).
\end{equation}

\section{Results}
\subsection{Single-mode matching}
The case of pump matching the single Schmidt mode of seed is described in \cite{brecht2011quantum, eckstein2011quantum}, where the possibility to block the selected mode has been demonstrated both theoretically and experimentally. Photon distribution in the non-matched Schmidt modes is not changed by the quantum optical gate. In its turn, photons in the matched Schmidt mode are being partly (or completely) converted into the sum-frequency mode, according to equations \eqref{eq:7} and \eqref{eq:8}. Here we will stress on some further important features found using our theoretical model.

Since there is a conversion between the signal and sum-frequency modes, one can be interested in the correlations between photons in these channels. The value of noise reduction factor between signal and sum-frequency modes is given by:
\begin{equation}\label{eq:14}
\mathrm{NRF} = \frac{\Delta[\widehat{N}_\text{s}^\text{in}] \cos^2 2\Theta + N_\text{s}^\text{in}  \sin^2 2\Theta}{ N_\text{s}^\text{in}} \geq 1,
\end{equation}
and its minimum is equal to unit and is achieved at beamsplitter angle $\Theta = \pi/4$, 
hence no correlation between signal and output photons can be observed.

As for squeezing in matched Schmidt mode, calculation of the variances of quadratures shows that it is still in squeezed state:
\begin{eqnarray}
& \Delta\left[\widehat{X}_n^\text{out}\right] = \frac{1}{2}\cdot\left\{ \sin^2 \Theta + \cos^2 \Theta \cdot e^{+2G\sqrt{\lambda_n}} \right\}, \label{eq:15} \\
& \Delta\left[\widehat{P}_n^\text{out}\right] = \frac{1}{2}\cdot\left\{ \sin^2 \Theta + \cos^2 \Theta \cdot e^{-2G\sqrt{\lambda_n}} \right\}. \label{eq:16}
\end{eqnarray}
For angles of full conversion these variances correspond to the vacuum, which indicates that all photons have been up-converted. If the transformation of beamsplitter is trivial, that is for $\Theta = 2\pi n$, these variances correspond to the initial squeezed vacuum in these modes. For intermediate values of rotation angle $\Theta$ one of variances appears to be less than vacuum noise, whereas another exceeds it, though their product is found to be always larger than standard quantum limit $1/4$:
\begin{equation}\label{eq:17}
\Delta[\widehat{X}_n^\text{out}] \cdot \Delta[\widehat{P}_n^\text{out}] =\frac{1}{4}\left(1+ \sinh^2\left(G\sqrt{\lambda_n}/2\right) \cdot \sin^22\Theta\right).
\end{equation}
Thus the degree of squeezing decreases after SFG process, but still one of variances remains suppressed.

The calculation of variances of quadratures of light generated in sum-frequency mode shows that it may be squeezed as well:
\begin{eqnarray}
& \Delta \left[\widehat{X}_\text{SF}^\text{out}\right] = \frac{1}{2}\cdot\left\{ \cos^2 \Theta + \sin^2 \Theta \cdot e^{+2G\sqrt{\lambda_n}} \right\}, \label{eq:18} \\
& \Delta \left[\widehat{P}_\text{SF}^\text{out}\right] = \frac{1}{2}\cdot\left\{ \cos^2 \Theta + \sin^2 \Theta \cdot e^{-2G\sqrt{\lambda_n}} \right\}. \label{eq:19}
\end{eqnarray}
Note that for full conversion angles these variances correspond exactly to the squeezed vacuum state. In other words, squeezed vacuum light of signal Schmidt mode is fully ``transferred'' to the sum-frequency mode, which is spectrally distributed according to \eqref{eq:4}. Such operation illustrates the application of quantum optical gate as the frequency converter for squeezed vacuum light.

\begin{figure}[b!]
	\centering
	\includegraphics[width=1\linewidth]{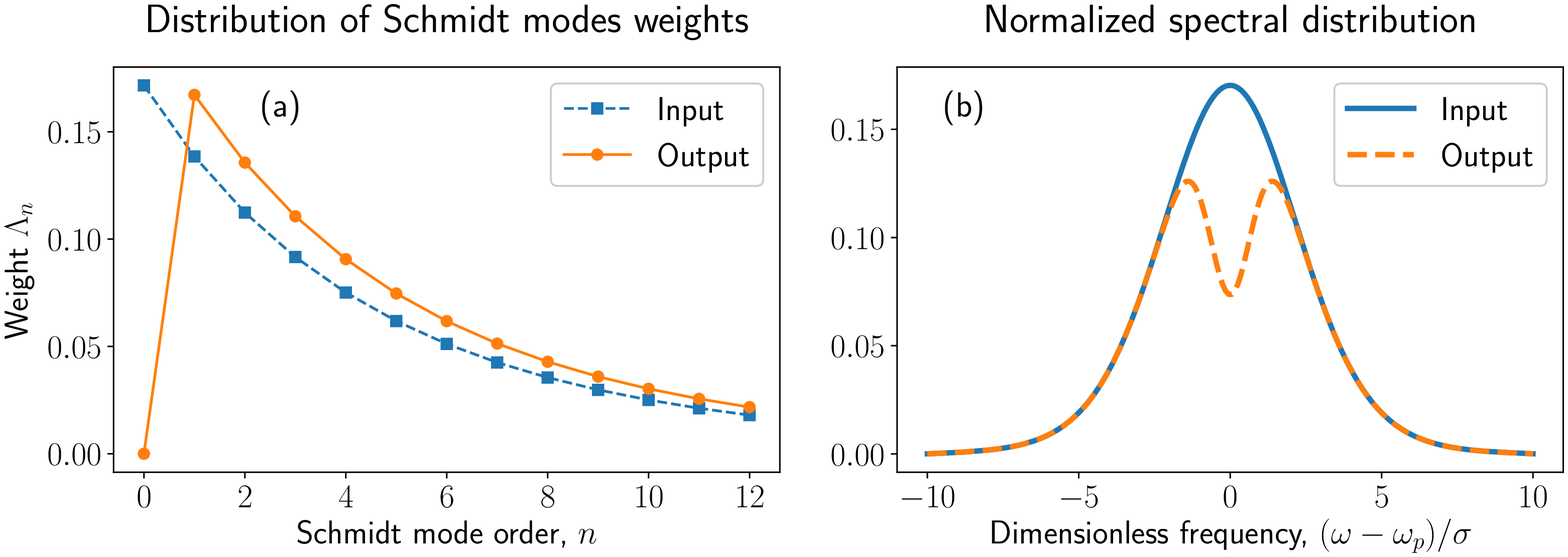}
	\caption{Blocking of zero-order Schmidt mode of squeezed vacuum from the seed with squeezing parameter $G\sqrt{\lambda_0}  = 4.39$. (a) Redistribution of weights of Schmidt model; (b) Corresponding change of spectral distribution, divided by input number of photons.}
	\label{fig:1}
\end{figure}

At the same time full conversion ($\Theta = \pi/2$) leads to strong redistribution of weights of Schmidt modes in squeezed vacuum seed and therefore changes significantly the spectral density of passed squeezed light. Fig.\ref{fig:1} demonstrates the regime of blocking the zero-order Schmidt mode and consequent redistribution of the Schmidt mode weights with predominant contribution of the mode with $n = 1$. The corresponding spectral signal is changed dramatically and reveals significant minimum in the center instead of initial maximum due to presence of the zero order Schmidt mode.

Another possible application of full conversion regime apart opposite to the blocking the seeding Schmidt mode is the selection of certain Schmidt mode of incoming squeezed light. For this purpose, two sequent SFG gates are required. The first one is seeded by squeezed vacuum light and a certain Schmidt mode is fully converted into the sum-frequency mode, which appears to be in a squeezed vacuum state according to \eqref{eq:18}-\eqref{eq:19}. Generated signal from $\widehat{C}$-mode is further utilized as a seed for the sum-frequency channel of second SFG unit, while no seed is used for the signal channel. If second SFG-gate is tuned to the full conversion as well, the photons from the sum-frequency seed will be down-converted to the signal mode. Ultimately, the squeezed light only in one selected Schmidt mode is generated in a second SFG unit.

\subsection{Two-mode matching}
In comparison to the results of the previous section, schemes with signal mode seeded by the superposition of several Schmidt modes seem to be very promising as well. In this case the signal mode $\Phi(\omega)$ is expanded over a set of Schmidt modes $\Phi(\omega) = \sum_{n} \mu_n \cdot u_n(\omega)$ with decomposition coefficients satisfying the normalization condition $\sum_{n}|\mu_n|^2 = 1$. Photons in the Schmidt modes with non-zero projections are involved in the conversion by SFG-gate, since the signal mode operator \eqref{eq:5} can be represented as the following superposition of Schmidt operators:
\begin{equation}\label{eq:20}
\widehat{D} = \sum_{n=0}^\infty \mu_n \widehat{A}_n.
\end{equation}
Note that in general the coefficients $\mu_n$ are complex with phases determined by both signal mode and involved Schmidt modes. By the substitution of \eqref{eq:19} into the SFG Hamiltonian \eqref{eq:3}, we get the solution of the Heisenberg equations for the Schmidt operators which is given by:
\begin{equation}\label{eq:21}
\widehat{A}_n^\text{out} = \widehat{A}_n^\text{in} - \mu_n^* \sin \Theta \cdot \widehat{C}^\text{in} + \mu_n^* \left( \cos \Theta - 1 \right) \cdot  \widehat{D}^\text{in} .
\end{equation}
This equation plays a fundamental role in the further analysis, as it gives the evolution of each individual Schmidt mode depending on its overlap with signal mode. The structure of \eqref{eq:21} implies strong interplay between initially independent Schmidt modes, since even partial conversion of one mode influences the solutions for others. Further we assume that the signal mode is the superposition of only two Schmidt modes of order $n_1, n_2$, and for brevity the corresponding projections are denoted as $\mu_1$ and $\mu_2$ respectively. 

\begin{figure}[b!]
	\centering
	\includegraphics[width=1	\linewidth]{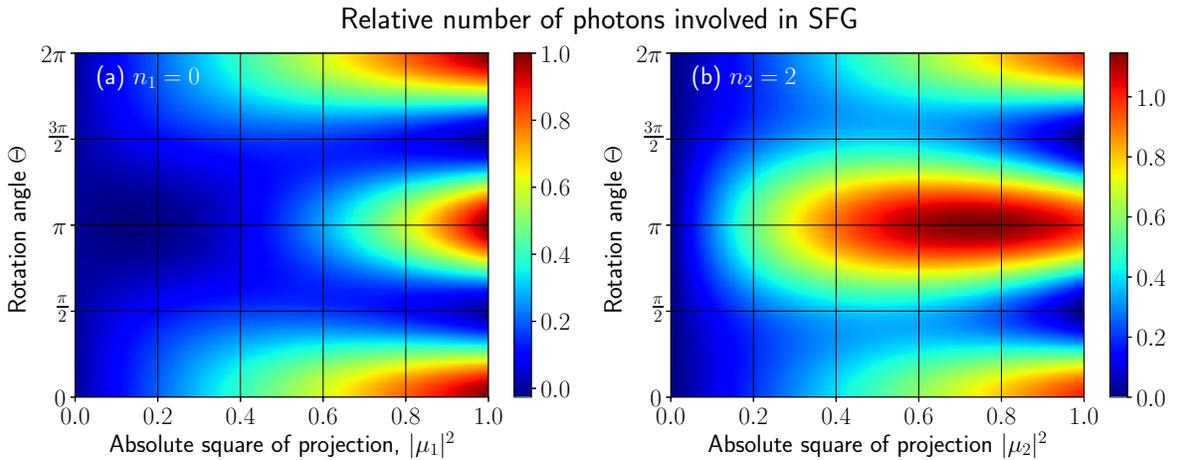}
	\caption{The dependence of relative number of photons involved in SFG, which is calculated according to \eqref{eq:30} without the first term and then divided by the input number of photons in corresponding mode, in the case of matching modes with $n_1 = 0$ (a) and $n_2 = 2$ (b) of input squeezed vacuum with $G\sqrt{\lambda_0} = 4.39$.}
	\label{fig:2}
\end{figure}

Similarly to the beasmplitter matrix \eqref{eq:6} which is simply the rotation on the angle $\Theta$, the transformation \eqref{eq:21} is found also to describe a rotation in the subspace of considered operators (details are in appendix B):
\begin{equation}\label{eq:22}
\begin{pmatrix}
\widehat{C}^\text{out} \\
\widehat{A}_1^\text{out} \\
\widehat{A}_2^\text{out} 
\end{pmatrix}
=  
\begin{pmatrix}
\cos{\Theta} & \mu_1   \sin{\Theta} & \mu_2   \sin\Theta  \\
-\mu_1^* \sin{\Theta}  & \cos\Theta+|\mu_2|^2 (1-\cos{\Theta}) & \mu_1^* \mu_2 (\cos\Theta - 1) \\
-\mu_2^* \sin\Theta & \mu_1 \mu_2^* (\cos \Theta - 1) & \cos \Theta + |\mu_1|^2 (1-\cos \Theta)
\end{pmatrix} \cdot \begin{pmatrix}
\widehat{C}^\text{in} \\
\widehat{A}_1^\text{in} \\
\widehat{A}_2^\text{in} 
\end{pmatrix}.
\end{equation}
The conservation law \eqref{eq:9} of photons in signal and sum-frequency modes
\begin{equation}\label{eq:23}
N_s(t) + N_\text{SF}(t) = N_s^\text{in} = |\mu_1|^2 \cdot N_1^\text{in} + |\mu_2|^2\cdot N_2^\text{in},
\end{equation}
wherein the number of photons in signal and sum-frequency modes:
\begin{eqnarray}
& N_s^\text{out} = \cos^2\Theta \cdot \sum_{k=1,2} |\mu_k|^2 N_k^\text{in},				\label{eq:24}\\
& N_\text{SF}^\text{out} = \sin^2 \Theta \cdot \sum_{k=1,2} |\mu_k|^2 N_k^\text{in},	\label{eq:25}
\end{eqnarray}	
with $N_1^\text{in}, N_2^\text{in}$ calculated according to \eqref{eq:13}. The result \eqref{eq:23} indicates that not all photons in matched modes are involved in the transformation by SFG-gate. The remaining part of photons can be characterized by the broadband operator $\widehat{R}$ and spectral mode $\Phi_R(\omega)$ orthogonal to the signal mode $\Phi(\omega)$, which is easily found to be	
\begin{equation}\label{eq:26}
\Phi_R(\omega) = - \mu_2^* \cdot u_1(\omega) + \mu_1^* \cdot u_2(\omega),
\end{equation}	
with the corresponding operator responsible for creation and annihilation of photons in this mode given by:	
\begin{equation}\label{eq:27}
\widehat{R} = - \mu_2^* \widehat{A}_1 + \mu_1^* \widehat{A}_2.
\end{equation}	
It can be immediately established that this operator as well as corresponding operator of photon number in mode \eqref{eq:26} are the integral of motion since they commute with the Hamiltonian \eqref{eq:3}. Conservation of the last quantity leads to the additional integral of motion, which is generalization of basic photon conservation law for the multimode case:
\begin{equation}\label{eq:28}
\widehat{N}_1 (t) + \widehat{N}_2(t) - \widehat{N}_s(t) = \widehat{N}_R^\text{in},
\end{equation}
with mean number $N_R^\text{in}$
\begin{equation}\label{eq:29}
N_R^\text{in} = \sum_{k=1,2} \left( 1 - |\mu_k|^2 \right) N_k^\text{in}.
\end{equation}
Using the transformation \eqref{eq:22} we may calculate the mean number of photons in all channels involved in gate transformation. The photon output in matched Schmidt modes is found to be
\begin{equation}\label{eq:30}
N_i^\text{out}    = \left(1-|\mu_i|^2 \right) N_i^\text{in} + \left(|\mu_i|^2 \cos^2\Theta \right) N_i^\text{in}  - |\mu_i\mu_j|^2 (\cos \Theta - 1)^2 \left( N_i^\text{in} - N_j^\text{in} \right),
\end{equation}
with $i = 1,2$ and $j=2,1$ respectively. The last term in \eqref{eq:30} shows the exchange of photons between two Schmidt modes, which is illustrated on Fig.\ref{fig:2} for seeding modes with $n_1 = 0$ and $n_2 = 2$.
Further we analyze different regimes and illustrate possible application of considered scheme and assume for simplicity that projections of Schmidt modes on the signal mode have equal absolute values $|\mu_1| = |\mu_2| = 1/\sqrt{2}$. In single-mode matching case the full up-conversion corresponds to the angle of beamsplitter $\Theta = \pi/2$. In multimode matching, however, the full conversion is not feasible, and rotation on such angle results in equal number of photons in both matched modes after the SFG-gate:
\begin{equation}\label{eq:31}
N^\text{out}_1  =  N^\text{out}_2  = \frac{N_1^\text{in} + N_2^\text{in}}{4},
\end{equation}
with the rest of input signal photons up-converted into the sum-frequency mode. In this case the involved Schmidt modes in SFG-gate cease to be independent and the output photon operators in these modes appear to be connected with each other:
\begin{equation}\label{eq:32}
\widehat{C}^\text{out} = - \dfrac{\widehat{A}_1^\text{out} + \widehat{A}_2^\text{out}}{\sqrt{2}}, 
\end{equation}
which is found for $\mu_1 = \mu_2 = 1/\sqrt{2}$.

\begin{figure}[b!]
	\centering
	\includegraphics[width=1\linewidth]{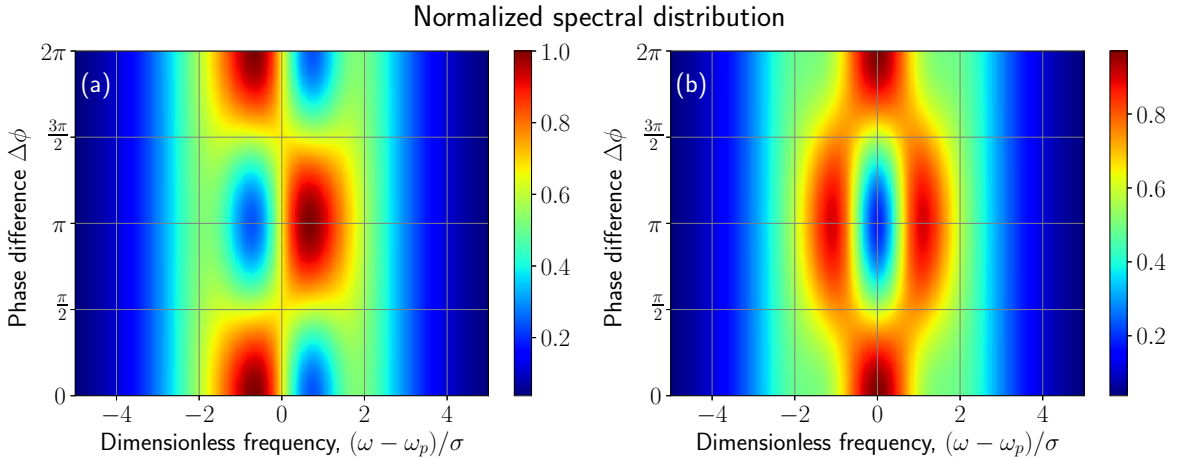}
	\caption{Spectral distribution of outcoming squeezed light ($G\sqrt{\lambda_0} = 6.59$) calculated as the function of phase difference between matched modes in regime of equal absolute squares of projections with $\Theta = \pi/2$. (a) Matching zero-order $n_1 = 0$ and first-order $n_2 = 1$ modes, (b) matching $n_1 = 0$ and $n_2 = 2$. Spectrum is normalized to a maximum of distribution.}
	\label{fig:3}
\end{figure}

Another consequence of interplay between matched Schmidt mode is the appearance of interference term in the spectral distribution of squeezed vacuum \eqref{eq:11}. Since the Schmidt modes can be no longer independent, in addition to the result \eqref{eq:12} the following interference part is found to contribute to the spectral intensity:
\begin{equation}\label{eq:33}
I(\omega) = 2  \cdot \mathrm{Re}\,\{ \mu_1^ * u^*_1(\omega) \cdot \mu_2 u_2(\omega) \}  (\cos \Theta - 1) \sum_{k = 1,2} \left(1 + |\mu_k|^2(\cos\Theta -1)\right) N_k^\text{in}.
\end{equation}
It can be seen that this interference term is crucially sensitive to phases of incoming Schmidt modes and their projections. This fact is illustrated by Fig.\ref{fig:3} where 2D plot of the total output intensity is presented in dependence on the frequency and the total phase $\Delta \phi$. This total phase is determined by the phases of projections $\mu_1$ and $\mu_2$ as well as by the constant phases of the Schmidt modes (such as $i^n$) while their frequency-dependent phases mostly cancel each other. Different intensity profiles can be obtained by varying the total phase. Thus it is possible to control the output distribution by changing the phase and vice versa to extract the information about the phase from the spectrum. Note, that the data of Fig.\ref{fig:3} correspond to the beamsplitter angle $\Theta = \pi/2$ when the role of the interference term is mostly pronounced due to strong influence of the Schmidt modes on each other. The interference contribution totally vanishes for $\Theta = \pi n$ when the modes are independent.

\begin{figure}[b!]
	\centering
	\includegraphics[width=1\linewidth]{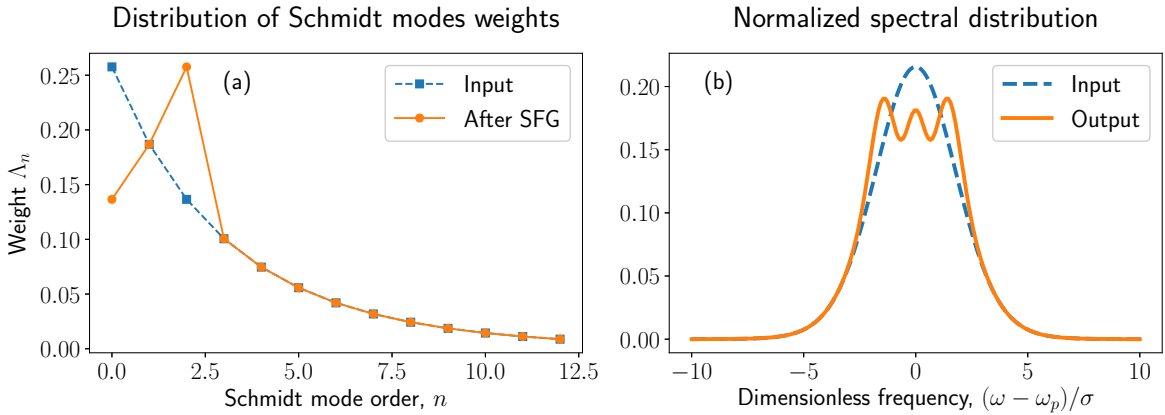}
	\caption{Swapping the weights of zero and second order Schmidt modes of squeezed light seed with $G\sqrt{\lambda_0} = 6.59$. (a) Redistributed weights of Schmidt modes; (b) corresponding change of spectral signal divided by input number of photons.}
	\label{fig:4}
\end{figure}

There is also a possibility to provide the complete exchange of photons between matched modes without any conversion to sum-frequency, i.e. $N_1^\text{out} = N_2^\text{in}$ and $N_2^\text{out} = N_1^\text{in}$. Such operation is achieved by the rotation on angle $\Theta = \pi$, which in single-mode gate results in trivial transformation, however in multimode matched gate such regime leads to significant redistribution of mode weights and change of spectral signal (see Fig.\ref{fig:4}). Moreover, transformed operators of matched Schmidt modes acquire the phase shift, which leads to the rotation of the axis of squeezing of quadratures. By introducing $\mu_i = |\mu_i| \cdot e^{i\phi_i}$, we obtain:
\begin{eqnarray}
& \widehat{A}_1^\text{out} = - \widehat{A}_2^\text{in} \cdot e^{+i( \phi_2 - \phi_1) }, \label{eq:34}\\
& \widehat{A}_2^\text{out} = - \widehat{A}_1^\text{in} \cdot e^{-i( \phi_2 - \phi_1) }. \label{eq:35}
\end{eqnarray}

\subsection{Twin-beam seeding}
Throughout previous analysis the squeezed light seed for SFG process was assumed to be generated in the degenerate and collinear regime of parametric down-conversion process. However, there is regime when PDC process yields the twin-beams of squeezed light, which exhibit strong correlations between of signal and idler photons. The corresponding state vector of two-mode squeezed light is given by \cite{Sharapova:2018}:
\begin{equation}\label{eq:36}
|\psi\rangle_s = \exp{\left\{G\sum_{n=0}^{\infty} \sqrt{\lambda_n}\left( \widehat{A}_n^\dagger \widehat{B}_n^\dagger - \widehat{A}_n \widehat{B}_n\right) \right\}} |0\rangle_s,
\end{equation}
where Schmidt operators $\widehat{B}_n$ are responsible for the idler photons operators. Signal and idler photons in the same order are correlated in that sense that variance of photon number difference is zero if both modes are in vacuum state before PDC
\begin{equation}\label{eq:37}
\Delta[\widehat{N}^\text{in}_\text{diff}] = 
\Delta[\widehat{A}_n^{\text{in }\dagger} \widehat{A}^{\text{in}}_n - \widehat{B}_n^{\text{in } \dagger} \widehat{B}^{\text{in}}_n] = 0,
\end{equation}
in other words, detection of signal photon is accompanied by the detection of idler photon, and vice versa.

If one channel (for instance signal photons) of two-mode squeezed vacuum light is engaged in sum-frequency generation process, while the other one is not affected by it, it is possible to alter the correlation properties of initial twin-beam. In case when SFG-gate matches the single $n$-th order Schmidt mode the number of photons in this mode is changed and the variance of difference of photon numbers \eqref{eq:31} at the output of the SFG gate appears to depend on rotation angle:
\begin{equation}\label{eq:38}
\Delta[\widehat{N}^\text{out}_\text{diff}]   = \sin^2 \Theta \cdot \sinh^2(G\sqrt{\lambda_n})   \left( \cos^2 \Theta + \sin^2 \Theta \cdot \cosh^2(G\sqrt{\lambda_n})\right) ,
\end{equation}
where the expression on the right-hand side coincides with the variance of number of photons in sum-frequency channel $\Delta[\widehat{N}_\text{SF}]$. In other words, generation of sum-frequency photons ruins the correlation between signal and idler beams in selected mode.

However, in the two-mode matching regime at $\Theta = \pi$ there is no up-conversion of signal photons. Thus we suggest that SFG-gate is seeded by signal photons in two Schmidt modes with equal squares of projections $|\mu_1|^2 = |\mu_2|^2 = 1/2$ and tuned to rotation angle $\Theta = \pi$. The corresponding idler channels are not involved in the gate. Before the transformation the correlations \eqref{eq:37} are present between photons in the same order. However, after the swapping the weights of Schmidt modes involved in SFG-gate the correlations are found between the idler and swapped modes:
\begin{eqnarray}
& \Delta[\widehat{B}^{\text{out }\dagger}_1 \widehat{B}_1^{\text{out}}  - \widehat{A}_2^{\text{out }\dagger} \widehat{A}_2^{\text{out}}] = 0, \label{eq:39}\\
& \Delta[\widehat{B}_2^{\text{out }\dagger} \widehat{B}_2^{\text{out}} - \widehat{A}_1^{\text{out }\dagger} \widehat{A}_1^{\text{out}}] = 0. \label{eq:40}
\end{eqnarray}
It is important that now the correlations are observed between photons in unchanged idler mode and new swapped signal Schmidt mode. Thus it is possible to alter the spectral distribution of correlated photons which can be useful for practical applications and purposes. 

\section{Conclusions}
In this paper we describe theoretically the capabilities of quantum optical gate for non-classical squeezed light which is based on the sum-frequency generation process seeded by several Schmidt modes of squeezed light. We focus mostly on the multi-mode regime when several modes of the seeding field are involved in the gate instead of the only one. In this case the interplay between initially independent Schmidt modes becomes possible, which allows to manage and manipulate extensively the properties and the mode content of incoming non-classical light. The most important feature of the considered scheme is significant sensitivity to the phases of involved Schmidt modes and therefore their projections on the signal mode profile. Different regimes of the SFG gate are examined and the possibilities to control the coupling between modes and to tailor the spectral distribution of the output light are demonstrated. Methods to change or choose the spectral mode of photons correlated with the idler beam in the case of twin-beam seeding are suggested. The effect of swapping between modes involved in the gate is found which allows one to enhance the output signal in a set of modes with required frequency distribution and without loss of photon correlations or squeezing. The obtained results seem to be very useful for further experiments and new applications with non-classical light. 

\section*{Funding Information}
We acknowledge financial support of the joint DFG-Russian Science Foundation (RSF) project SH 1228/2-1, ME 1916/7-1 - No.19-42-04105.

\section*{Acknowledgments}
We gratefully thank Prof. Christine Silberhorn for fruitful discussions.

\appendix
\numberwithin{equation}{section}

\newpage 
\section*{Appendix}
\section{Derivation of Hamiltonian}\label{sec:Hamiltonian}
The pump field is classical and has the normalized spectral distribution $\Phi(\omega)$ and amplitude $E_0$, whereas for quantized fields of signal and output modes one applies the expansion over plane-waves:
\begin{eqnarray}
&& \widehat{E}_{s}^{(+)} (\mathbf{r},t) =  \int d\mathbf{k}_{s}\, C_{\mathbf{k}_{s}}\, e^{i\left(\mathbf{k}_{s} \mathbf{r} - \omega_s t\right)} \, \widehat{a}_{\mathbf{k}_{s}}, \label{eq:A1}\\
&& \widehat{E}_{o}^{(-)} (\mathbf{r},t) =  \int d\mathbf{k}_{o}\, C_{\mathbf{k}_{o}}\, e^{-i\left(\mathbf{k}_{o} \mathbf{r} - \omega_o t\right)} \, \widehat{a}^\dagger_{\mathbf{k}_{o}}. \label{eq:A2}
\end{eqnarray}

As only the collinear propagation is taken into account, one may replace the integration over wave-vector components in \eqref{eq:A1} and \eqref{eq:A2} by the integration over frequencies, with the assumption of zero energy mismatch: absorption of pump photon with energy $\hbar \omega$ and signal photon $\hbar \omega_s$ results in the generation of sum-frequency photon with energy $\hbar \omega_o = \hbar \omega + \hbar\omega_s$. Then after the integration over spatial coordinate of crystal along $z$-axis Hamiltonian \eqref{eq:1} becomes:
\begin{equation}\label{eq:A3}
\widehat{H}_\text{SFG} = i\hbar \Gamma \int d\omega_s d\omega_o \, F(\omega_s, \omega_o)  \widehat{a}_{\omega_s}  \widehat{a}^\dagger_{\omega_o} + h.c.,
\end{equation}
where $F(\omega_s, \omega_o)$ is the normalized two-photon amplitude of SFG process, which is a product of pump spectral envelope and the phase-matching function:
\begin{eqnarray}\label{eq:A4}
F(\omega_s, \omega_o)  \sim \Phi(\omega_s + \omega_p - \omega_o)\,
\exp\left\{\frac{i\Delta k_z L}{2}\right\}\, \mathrm{sinc}\left(\frac{\Delta k_z L}{2}\right).
\end{eqnarray}
where $\Delta k_z$ is the wave-vector mismatch inside the crystal:
\begin{equation}\label{eq:A5}
\Delta k_z = k_p(\omega_p) + k_s(\omega_s) - k_o(\omega_o).
\end{equation}
In the first order of Taylor expansion we have:
\begin{equation}\label{eq:A6}
\Delta k_z \approx \left[k_p'(\omega_p) - k_o'(2\omega_p) \right] \cdot (\omega_o - 2\omega_p),
\end{equation}
where the zero order vanishes due to crystal tuning $k_p(\omega_p) + k_i(\omega_p) - k_o(2\omega_p) = 0$, i.e. exact matching is achieved for the centers of envelopes. Upon the approximation of sinc function by the Gaussian \cite{Fedorov:2009}, the two-photon amplitude \eqref{eq:A4} becomes:
\begin{eqnarray}\label{eq:A7}
F(\omega_s, \omega_o) \sim\Phi(\omega_s + \omega_p - \omega_o)\cdot \exp\left[-\frac{(\omega_o - 2\omega_p)^2}{2\Delta\omega^2}\right],
\end{eqnarray}
where $1/\Delta \omega \sim L\cdot |k_p'(\omega_p) - k_o'(2\omega_p)|$. If the pump duration satisfies the condition \eqref{eq:2}, the frequency of output photon $\omega_o$ is localized in the small proximity of $2\omega_p$, which allows the spectral factorization of two-photon amplitude, i.e. $\Phi\big(\omega_s - \omega_p - (\omega_o - 2\omega_p)\big) \approx \Phi(\omega_s - \omega_p)$, and we get
\begin{equation}\label{eq:A9}
F(\omega_s, \omega_o)  \approx  \Phi(\omega_s - \omega_p) \cdot f(\omega_o - 2\omega_p) .
\end{equation}
where $f(\omega_o - 2\omega_{p})$ is the phase-matching function which describes the wave-packet of sum-frequency:
\begin{equation}\label{eq:A10}
f(\omega_o - 2\omega_p)  =  \sqrt[4]{\frac{1}{\pi \Delta\omega^2}} \exp\left[-\frac{(\omega_o - 2\omega_p)^2}{2\Delta\omega^2}\right].
\end{equation}

Substitution of \eqref{eq:A9} and \eqref{eq:A10} into the Hamiltonian \eqref{eq:A3} results in the expression \eqref{eq:3} and  definition of new broadband modes \eqref{eq:4} and \eqref{eq:5}.

\section{Geometrical analysis}\label{sec:geometrical-analysis}
The solution of the Heisenberg equations of motion for operators $\widehat{C}$ \eqref{eq:6}, $\widehat{D}$ \eqref{eq:20} and $\widehat{R}$ \eqref{eq:27} can be given by the following matrix transformation:
\begin{equation}
\begin{pmatrix}
\widehat{C}^\text{out} \\
\widehat{D}^\text{out} \\
\widehat{R}^\text{out} 
\end{pmatrix}
=   \begin{pmatrix}
\cos \Theta & \sin \Theta & 0 \\
-\sin\Theta & \cos \Theta & 0  \\
0& 0   & 1
\end{pmatrix}  \cdot \begin{pmatrix}
\widehat{C}^\text{in} \\
\widehat{D}^\text{in} \\
\widehat{R}^\text{in} 
\end{pmatrix}.
\end{equation}
which is the rotation at the angle $\Theta$ around axis determined by conserved operator $\widehat{R}$. 

At the same time, all these operators at the input of SFG-gate are connected with the Schmidt operators $\widehat{A}_1, \widehat{A}_2$ by unitary transformation, which is in its turn the rotation at some angle around the axis determined by $\widehat{C}$:
\begin{equation}\label{eq:B5}
\begin{pmatrix}
\widehat{C}^\text{in} \\
\widehat{D}^\text{in} \\
\widehat{R}^\text{in} 
\end{pmatrix}
=   \begin{pmatrix}
1 & 0 & 0 \\
0& \mu_1 & \mu_2  \\
0&-\mu_2^*  &\mu_1^*
\end{pmatrix}  \cdot \begin{pmatrix}
\widehat{C}^\text{in} \\
\widehat{A}_1^\text{in} \\
\widehat{A}_2^\text{in} 
\end{pmatrix}.
\end{equation}
The total evolution of the photon operators in the sum-frequency and seeding Schmidt modes provided by the SFG gate is given by:	
\begin{equation}
\begin{pmatrix}
\widehat{C}^\text{out} \\
\widehat{A}_1^\text{out} \\
\widehat{A}_2^\text{out} 
\end{pmatrix} =
\begin{pmatrix}
1 & 0 & 0 \\
0& \mu_1^* & -\mu_2  \\
0&\mu_2^*  &\mu_1
\end{pmatrix}  \cdot   \begin{pmatrix}
\cos \Theta & \sin \Theta & 0 \\
-\sin\Theta & \cos \Theta & 0  \\
0& 0   & 1
\end{pmatrix}  \cdot \begin{pmatrix}
1 & 0 & 0 \\
0& \mu_1 & \mu_2  \\
0&-\mu_2^*  &\mu_1^*
\end{pmatrix}\cdot \begin{pmatrix}
\widehat{C}^\text{in} \\
\widehat{A}_1^\text{in} \\
\widehat{A}_2^\text{in} 
\end{pmatrix},
\end{equation}
and can be interpreted as the beamsplitter rotation at the angle $\Theta$ modified by unitary transformation, which is the rotation on different angle around different axis. In other words, the resulting transformation of operators is the composition of multiple rotations in different subspaces of involved operators.

\newpage

\bibliographystyle{unsrt} 
\bibliography{sample} 

\end{document}